%Paper: hep-th/9406052
%From: LESNIEWSKI@huhept.harvard.edu
%Date: Wed, 8 Jun 1994 21:55:12 -0400 (EDT)

\newbox\leftpage \newdimen\fullhsize \newdimen\hstitle \newdimen\hsbody
\tolerance=10000\hfuzz=10000pt
%    tolerance=1000 \hfuzz=20pt
\def\printertype{ps: }% default postscript
\def\qms{\def\printertype{qms: }% qms lasergrafix
\ifx\answ\bigans\else\voffset=-.4truein\hoffset=.125truein\fi}
\def\bigans{b }
\message{ big or little (b/l)? }\read-1 to\answ
\ifx\answ\bigans\message{(This will come out unreduced.}
\magnification=1200\baselineskip=14pt plus 2pt minus 1pt
\hsbody=\hsize \hstitle=\hsize %take default values for unreduced format
\else\message{(This will be reduced.} \let\lr=L
\magnification=1000\baselineskip=16pt plus 2pt minus 1pt
\voffset=-.31truein\vsize=7truein\hoffset=-.59truein% apple lw
\hstitle=8truein\hsbody=4.75truein\fullhsize=10truein\hsize=\hsbody
\output={\ifnum\pageno=0 %%% This is the HUTP version
  \shipout\vbox{\special{\printertype landscape}\makeheadline
    \hbox to \fullhsize{\hfill\pagebody\hfill}}\advancepageno
  \else
  \almostshipout{\leftline{\vbox{\pagebody\makefootline}}}\advancepageno
  \fi}
\def\almostshipout#1{\if L\lr \count1=1 \message{[\the\count0.\the\count1]}
      \global\setbox\leftpage=#1 \global\let\lr=R
  \else \count1=2
    \shipout\vbox{\special{\printertype landscape}
      \hbox to\fullhsize{\box\leftpage\hfil#1}}  \global\let\lr=L\fi}
\fi

\input mssymb.tex
\font\bigfont=cmr10 scaled\magstep3

\def\section#1#2{\vskip32pt plus4pt \goodbreak \noindent{\bf#1. #2}
	\xdef\currentsec{#1} \global\eqnum=0 \global\thmnum=0}

\newcount\thmnum
\global\thmnum=0
\def\prop#1#2{\global\advance\thmnum by 1
	\xdef#1{Proposition \currentsec.\the\thmnum}
	\bigbreak\noindent{\bf Proposition \currentsec.\the\thmnum.}
	{\it#2} }
\def\define#1#2{\global\advance\thmnum by 1
	\xdef#1{Definition \currentsec.\the\thmnum}
	\bigbreak\noindent{\bf Definition \currentsec.\the\thmnum.}
	{\it#2} }
\def\lemma#1#2{\global\advance\thmnum by 1
	\xdef#1{Lemma \currentsec.\the\thmnum}
	\bigbreak\noindent{\bf Lemma \currentsec.\the\thmnum.}
	{\it#2}}
\def\thm#1#2{\global\advance\thmnum by 1
	\xdef#1{Theorem \currentsec.\the\thmnum}
	\bigbreak\noindent{\bf Theorem \currentsec.\the\thmnum.}
	{\it#2} }
\def\cor#1#2{\global\advance\thmnum by 1
	\xdef#1{Corollary \currentsec.\the\thmnum}
	\bigbreak\noindent{\bf Corollary \currentsec.\the\thmnum.}
	{\it#2} }

\newcount\eqnum
\global\eqnum=0
\def\num{\global\advance\eqnum by 1
	\eqno({\rm\currentsec}.\the\eqnum)}
\def\eqalignnum{\global\advance\eqnum by 1
	({\rm\currentsec}.\the\eqnum)}
\def\ref#1{\num  \xdef#1{(\currentsec.\the\eqnum)}}
\def\eqalignref#1{\eqalignnum  \xdef#1{(\currentsec.\the\eqnum)}}

\def\title#1{\centerline{\bf\bigfont#1}}

\newcount\subnum
\def\Alph#1{\ifcase#1\or A\or B\or C\or D\or E\or F\or G\or H\fi}
\def\subsec{\global\advance\subnum by 1
	\vskip12pt plus4pt \goodbreak \noindent
	{\bf \currentsec.\Alph\subnum.}  }
\def\newsubsec{\global\subnum=1 \vskip6pt\noindent
	{\bf \currentsec.\Alph\subnum.}  }
\def\today{\ifcase\month\or January\or February\or March\or
	April\or May\or June\or July\or August\or September\or
	October\or November\or December\fi\space\number\day,
	\number\year}

\def\intsec{I}
\def\chernsec{II}
\def\translsec{III}
\def\pathintsec{IV}
\def\bC{\Bbb{C}}
\def\bR{\Bbb{R}}
\def\bZ{\Bbb{Z}}
\def\h{{\cal H}}
\def\a{{\cal A}}

\def\Tr{\mathop{\rm Tr}\nolimits}

{\baselineskip=12pt
\nopagenumbers
\line{\hfill \bf HUTMP B332}
\line{\hfill \bf \today}
\vfill
\title{Superspace Formulation of the Chern Character}
\medskip
\title{of a Theta - Summable Fredholm Module}
\vskip1in
\centerline{{\bf Andrzej Lesniewski}$^*$\footnote{$^1$}
{ Supported in part by the Department of Energy under grant
DE--FG02--88ER25065} and {\bf Konrad Osterwalder}$^{**}$}
\vskip12pt
\centerline{ $^*$Lyman Laboratory of Physics}
\centerline{Harvard University}
\centerline{Cambridge, MA 02138}
\centerline{USA}
\vskip12pt
\centerline{ $^{**}$Department of Mathematics}
\centerline{ETH}
\centerline{8092 Z\"urich}
\centerline{Switzerland}
\vskip1in\noindent
{\bf Abstract.} We apply the concepts of superanalysis to present
an intrisically supersymmetric formulation of the Chern character
in entire cyclic cohomology. We show that the cocycle condition
is closely related to the invariance under supertranslations.
Using the formalism of superfields, we find a path integral
representation of the index of the generalized Dirac operator.
\vfill\eject}

\section\intsec{Introduction}
\newsubsec
The purpose of this note is to present an intrinsically supergeometric
formulation of the Chern character of [7] in entire cyclic cohomology
[3]. The concept of a Fredholm module is closely related to the
structure of supersymmetric quantum theory. The construction of the
Chern character associated with a $\Theta$--summable Fredholm module
presented in [7] uses in an essential way ideas adopted from
supersymmetric quantum field theory. The $n$--th component of the Chern
character is written as a certain finite temperature $(n+1)$--point
Schwinger function integrated over an $n$-simplex. The physical
interpretation of the closedness of the Chern character under Connes'
coboundary operator $\partial$ remained, however, unclear.

The construction of this paper is based on the simple observation that
a more natural form of the Chern character arises if the integrals
over simplexes are replaced by Berezin integrals over supersimplexes.
(A supersimplex is a superdomain whose base is an ordinary simplex.)
This makes the supersymmetric nature of the Chern character
transparent. We find that the Chern character is invariant under the
$(1|1)$--dimensional supergroup of translations in the time direction
and in one extra fermionic direction. It is the invariance under
supertranslations in the fermionic direction which is equivalent to
the closedness of the Chern character under $\partial$. Furthermore,
this form of the Chern character lends itself well to
a path integral representation. This should be a useful technical
tool in studying the topological properties of Fredholm modules
arising in quantum field theory. Our discussion of this idea has
a preliminary character and is based on a number of technical
assumptions whose validity needs to be established in each situation
separately.

\subsec
The paper is organized as follows. Section II contains the
the superspace formulation of the Chern character. In Section III,
we discuss the relation between the cocycle condition satisfied
by the Chern character and its supertranslation invariance. Finally,
in Section IV we discuss the conditions under which the Chern
character and index of the generalized Dirac operator have natural
superpath integral representations.

\subsec
There are several different definitions of a supermanifold available.
For the purpose of this paper, it would not matter which one is
used. For convenience, we adopt the Berezin - Leites - Kostant
definition, see e.g. [1]. A supermanifold $\cal X$ is thus a
locally trivial ringed space whose structure sheaf is isomorphic
to the tensor product of the sheaf of smooth functions on a manifold
$X$ (called the base of $\cal X$), and a Grassmann algebra
$\bigwedge (E)$ over a vector space $E$. We denote the superalgebra
of global sections of this sheaf by $C^\infty({\cal X})$, and refer
to a set of (even and odd) generators of this superalgebra as the
coordinates of a ``point'' of $\cal X$. By $\widehat\otimes_\pi$ we
denote the $\bZ_2$-graded, completed projective tensor product of
sheaves. Structure sheaves of supermanifolds are sheaves of nuclear
vector spaces, and so using $\widehat\otimes_\pi$ is convenient as
it leads to natural functorial properties of cartesian products of
supermanifolds.

\medskip\noindent
{\bf Acknowledgement.} The work presented in this paper was done in
January 93 while the first named author was visiting the
Forschungsinstitut f\"ur Mathematik at ETH Z\"urich whose hospitality
he would like to gratefully acknowledge.

\section\chernsec{The Chern Character}

\newsubsec
Let $\a$ be a unital $\bC^*$--algebra. Recall [2] that a
$\Theta$-summable Fredholm module over $\a$ is a triple $(\h, Q,\rho)$,
consisting of a $\bZ_2$--graded Hilbert space $\h$, a self-adjoint
operator $Q$ on $\h$ which is odd with respect to the $\bZ_2$-grading
of $\h$, and a grading preserving $*$--homomorphism $\rho$ of $\a$ to
the $\bC^*$--algebra ${\cal L}(\h)$ of bounded linear operators on
$\h$. For convenience of notation, we suppress $\rho$ in all formulas
below. We require that
\item{$\bullet$} the subalgebra $\a_1:=\{a\in\a:\;\Vert [Q,a]\Vert<
\infty\}$ is dense in $\a$;
\item{$\bullet$} for all $\beta >0$,
$$
\Tr(\exp(-\beta Q^2))<\infty\; .
$$

\noindent
Throughout this paper we will make the simplifying assumption that
$\a$ is trivially graded meaning that all elements of $\a$ are even.
By $\Gamma$ we denote the grading operator on $\h$. It satisfies the
properties $\Gamma^2=I$ and $\Gamma^*=\Gamma$. By $Q_+:=
P_-QP_+$, where $P_{\pm}:=\;{1\over 2}(I\pm\Gamma)$, we denote the
restriction of $Q$ to the even subspace of $\h$. Finally, by
$\Tr(A)$ we denote the trace of $A\in {\cal L}(\h)$.

A $\Theta$--summable Fredholm module defines a fundamental cohomology
class in the entire cyclic cohomology of $\a$, the Chern character [3].
In the representation of [7] and [5], the Chern character is a sequence
of $(n+1)$--linear functionals on $\a$, $\tau^\beta=\{\tau^\beta_{n}\}
_{n=0}^{\infty}$ constructed as follows. For $A_0,\ldots , A_n\in
{\cal L}(\h)$ we define
$$
F^\beta_{n}(A_0, A_1,\ldots ,A_{n}):=\beta^{-n/2}
\int_{\sigma^\beta_{n}}\Tr\;(A_0 A_1(t_1)
\ldots A_{n}(t_{n})\exp(-\beta Q^2))dt,
\ref{\thermavref}
$$
where $A(t):=\exp(-tQ^2)A\exp tQ^2$ is the operator $A$ ``at the
Euclidean time $t$'' [8], and where $\sigma^\beta_n:=\{(t_1,\ldots
\; ,t_n)\in \bR^n: \; 0<t_1< \ldots <t_n<\beta\}$ is the $n$-simplex
of length $\beta$. We then set for $a_0,\ldots ,a_n\in\a$
$$
\tau^\beta_{n}(a_0,a_1,\ldots ,a_{n}) := F^\beta_{n}(\Gamma^{n+1}a_0,
[Q,a_1],\ldots [Q,a_n]).\ref{\chernref}
$$
For the future reference, we also define the following sequence of
multilinear functionals:
$$
\tilde\tau^\beta_{n}(a_0,a_1,\ldots ,a_{n}) := F^\beta_{n}(\Gamma^na_0,
[Q,a_1],\ldots [Q,a_n]).\ref{\modchernref}
$$

\subsec
Let $\bR_+^{1|1}$ denote the $(1|1)$-dimensional superdomain whose
base is the set of positive real numbers. The coproduct morphism
$\Delta : {\cal C}^\infty(\bR_+^{1|1})\longrightarrow {\cal C}
^\infty(\bR_+^{1|1})\widehat\otimes_\pi {\cal C}^\infty(\bR_+^{1|1})$
given by
$$
(\Delta f)(s,\delta; t, \epsilon):=f(s+t+\delta\epsilon,
\delta+\epsilon),\num
$$
and the counit morphism $\iota : {\cal C}^\infty(\bR_+^{1|1})
\longrightarrow \bR$ given by
$$
\iota f := f(0,0)\num
$$
furnish $\bR_+^{1|1}$ with the structure of a supersemigroup.
Informally, we will use the product notation, $(s,\delta)\cdot(t,
\epsilon)=(s+t+\delta\epsilon,\delta+\epsilon)$.
\prop\repprop{The morphism ${\cal C}^\infty(\bR_+^{1|1})
\longrightarrow {\cal C}^\infty(\bR_+^{1|1})\widehat\otimes_\pi
{\cal L}(\h)$ given by
$$
(t,\epsilon)\longrightarrow T(t,\epsilon):=
\exp(-tQ^2+\epsilon Q)\num
$$
defines a representation of $\bR_+^{1|1}$ on $\h$.}
\medskip\noindent
{\it Proof.} We note that
$$
\eqalign{
\exp \delta Q\;\exp \epsilon Q
&=(1+\delta Q)(1+\epsilon Q)\cr
&=1+(\delta+\epsilon)Q-\delta\epsilon Q^2=
\exp(-\delta\epsilon Q^2+(\delta+\epsilon)Q),\cr}
$$
and so
$$
\eqalign{
T(s,\delta)T(t,\epsilon)&=\exp(-sQ^2+\delta Q)
\exp(-tQ^2+\epsilon Q)\cr
&=\exp(-(t+s+\delta\epsilon)Q^2+(\delta+\epsilon)Q)=
T((s,\delta)\cdot(t,\epsilon)),\cr}
$$
and the claim follows. $\square$

\subsec
For a positive integer $n$ and $\beta >0$, we define the
$(n|n)$--supersimplex $\sigma^\beta_{n|n}$ to be an
$(n|n)$--dimensional superdomain [1] whose base is the
$n$--simplex $\sigma^\beta_n$.
We let $t_1,\ldots ,t_n$ and $\epsilon_1,\ldots ,
\epsilon_n$ denote the even and odd generators of the structure sheaf
of $\sigma^\beta_{n|n}$, respectively. By $\int_{\sigma^\beta_{n|n}}
\ldots dt\, d\epsilon$ we denote the usual Berezin integral [1].

We set for $a\in\a$ and $(t,\epsilon)\in\sigma^\beta_{1|1}$,
$$
a(t,\epsilon):=\exp(-tQ^2+\epsilon Q)
a\exp(tQ^2-\epsilon Q).\num
$$
We note that $a(t)=a(t,0)$.
\prop\techprop{We have the following identities:
\item{(i)}
$$
\int a(t,\epsilon)d\epsilon = [Q,a](t);\ref{\epsintref}
$$
\item{(ii)} As operators on ${\rm Ran}(\exp(-\beta Q^2))$,
$$
\exp(-sQ^2+\delta Q)a(t,\epsilon)
\exp(sQ^2-\delta Q)= a((s,\delta)
\cdot(t,\epsilon)).\num
$$}
\medskip\noindent
{\it Proof.} The first identity follows from the observation that
$$
a(t,\epsilon)=a(t)+\epsilon\exp(-tQ^2)[Q,a]\exp tQ^2\; .
$$
The proof of the second identity is similar to the proof of \repprop\ .
$\square$
\subsec
An immediate consequence of \chernref\ and \techprop\ (i) is the
following representation of the Chern character by means of a
Berezin integral over the supersimplex $\sigma^\beta_{n|n}$,
$$
\tau^\beta_{n}(a_0,a_1,\ldots ,a_{n})=\beta^{-n/2}\int_{\sigma^\beta
_{n|n}}\Tr(\Gamma^{n+1}a_0a_1(t_1,\epsilon_1)\dots a_{n}(t_{n},
\epsilon_{n})\exp(-\beta Q^2))dt d\epsilon\; .\ref{\taudef}
$$
A similar representation holds for $\tilde\tau^\beta_n$.

\section\translsec{Translation Invariance in Superspace and
Connes' Coboundary}

\medskip
An immediate consequence of the representation \taudef ,
an analogous formula for $\tilde\tau^\beta_n$ and the cyclicity
of the trace is the following translation invariance property.
\prop\invprop{The functionals $\tau^\beta_{n}$ and $\tilde
\tau^\beta_{n}$are invariant under the supersemigroup $\bR_+^{1|1}$,
$$
\eqalign{
&\tau^\beta_{n}(a_0(s,\eta),a_1(s,\eta),\ldots ,a_{n}(s,\eta))=
\tau^\beta_{n}(a_0,a_1,\ldots ,a_{n}),\cr
&\tilde\tau^\beta_{n}(a_0(s,\eta),a_1(s,\eta),\ldots ,a_{n}(s,\eta))=
\tilde\tau^\beta_{n}(a_0,a_1,\ldots ,a_{n}).\cr
}\ref{\trinvref}
$$}
\medskip
Let us now take the left derivative $\partial/\partial\eta$ of
the second equation in \trinvref\ and evaluate it at $s = 0,\;
\eta = 0$. Keeping in mind the sign count in the super Leibniz
rule, we obtain the equation
$$
\sum_{0\leq j\leq n}\;(-1)^j\tilde\tau^{\beta}_n(a_0,\ldots ,
[Q,a_j],\ldots , a_n)=0.\num
$$
We claim that the above identity is just the cocycle condition
$\partial\tau^\beta = 0$, where $\partial = B + b$ is Connes'
coboundary operator of entire cyclic cohomology (see [2] for the
definitions of $b$ and $B$; below we use the normalization
of [7]). This follows from the proposition formulated below.
The proof of this proposition can be easily extracted from the
computations on pages 12--13 of [7]. For completeness, we present
a superspace version of these computations.

\prop\deltaprop{We have the identities
$$
(B\tau^\beta)_n(a_0,a_1,\ldots ,a_n) =
\beta^{1/2}\tilde\tau^\beta_n([Q,a_0], a_1,\ldots ,a_n),
\ref{\idoneref}
$$
and
$$
(b\tau^\beta)_n(a_0,a_1,\ldots ,a_n)=\beta^{1/2}
\sum_{1\leq j\leq n}\; (-1)^j\tilde\tau^{\beta}_n(a_0,
\ldots ,[Q,a_j],\ldots , a_n).
\ref{\idtworef}
$$}
\medskip\noindent
{\it Proof.} The proof of \idoneref\ is a straightforward
calculation. Using cyclicity of the trace, we find
$$
\eqalign{
&\beta^{(n+1)/2}(B\tau^{\beta})_n(a_0,a_1,\ldots ,a_n)=\cr
&\qquad\beta^{(n+1)/2}\sum_{j=0}^n(-1)^{nj}\tau^{\beta}_{n+1}
(I, a_{n+1-j},\ldots , a_n, a_0,\ldots , a_{n-j})=\cr
&\qquad\sum_{j=0}^n(-1)^{nj}\int_{\sigma^\beta_{n+1|n+1}}
\Tr (\Gamma^na_{n+1-j}(t_1,\epsilon_1)\ldots \cr
&\qquad\qquad\qquad a_n(t_{j-1},
\epsilon_{j-1})a_0(t_j,\epsilon_j)\ldots a_{n-j}(t_{n-j},
\epsilon_{n-j})e^{-\beta Q^2})dt d\epsilon =\cr
&\qquad\sum_{j=0}^n\int_{\sigma^\beta_{n+1|n+1}}
\Tr (\Gamma^na_0(0,\epsilon_0)\ldots a_j(t_j,\epsilon_j)
a_{j+1}(t_{j+2},
\epsilon_{j+1})\ldots a_n(t_{n+1},\epsilon_n)e^{-\beta Q^2})
dt d\epsilon\; .\cr
}
$$
Integrating out the redundant variables, we rewrite the above
expression as
$$
\eqalign{
&\sum_{j=0}^n\int_{\sigma^\beta_{n|n}}
(t_{j+1}-t_j)\Tr (\Gamma^n[Q,a_0]a_1(t_1,\epsilon_1)
\ldots a_n(t_n,\epsilon_n)\exp(-\beta Q^2)) dt d\epsilon =\cr
&\quad\beta\int_{\sigma^\beta_{n|n}}\Tr(\Gamma^n
[Q,a_0]a_1(t_1,\epsilon_1)\dots a_{n}(t_{n},
\epsilon_{n})\exp(-\beta Q^2))dt\; d\epsilon\; =\cr
&\quad\beta^{(n+2)/2}\tilde\tau^\beta_n([Q,a_0], a_1,
\ldots ,a_n),\cr
}
$$
where $t_0:=0, t_{n+1}:=\beta$, and our assertion is proven.

We now prove \idtworef . Using \techprop\ (ii), we obtain the
representation
$$
\eqalign{
&\beta^{n/2}\;\tilde\tau^\beta_n(a_0,a_1,\ldots , [Q, a_j],
\ldots ,a_n)=\cr\quad &\int\int_{\sigma^\beta_{n|n}}\Tr(\Gamma^n
a_0a_1(t_1,\epsilon_1)\dots a_j(t_j-\delta\epsilon_j, \epsilon_j
+\delta)\ldots a_{n}(t_n,\epsilon_n)\exp(-\beta Q^2))dt d\epsilon
d\delta\; .\cr}\ref{\interm}
$$
But $a_j(t_j-\delta\epsilon_j,\epsilon_j+\delta)=
a_j(t_j-\delta\epsilon_j,0)=a_j(t_j-\delta\epsilon_j)$, because
of $\epsilon_j^2=\delta^2=0$. At this point one is tempted to
make a change of variables $t_j-\delta\epsilon_j\rightarrow
t_j$ to conclude that the integral is zero. This, however,
is incorrect, as the integrand in \interm\ is not compactly
supported [1]. Instead, we have for $j = 1, 2, \ldots , n$,
$$
\int a_j(t_j-\delta\epsilon_j)dt_j d\epsilon_j d\delta =
-\int_{t_{j-1}}^{t_{j+1}}{d\over{dt_j}} a_j(t_j)\; dt_j=
a_j(t_{j-1})-a_j(t_{j+1}),\num
$$
(we set here $t_0:=0$, $t_{n+1}:=\beta$, and $\epsilon_0=
\epsilon_{n+1}:=0$) and consequently
$$
\eqalign{
&\beta^{n/2}\;\tilde\tau^\beta_{n}(a_0,a_1,\ldots , [Q, a_j],
\ldots , a_{n})=\cr
&-\int_{\sigma^\beta_{n-1|n-1}}\Tr(\Gamma^n
a_0a_1(t_1,\epsilon_1)\dots a_j(t_j)a_{j+1}(t_j,
\epsilon_j)\ldots a_n(t_n,\epsilon_n)e^{-\beta Q^2})dt
d\epsilon\; + \cr
&\int_{\sigma^\beta_{n-1|n-1}}\Tr(\Gamma^n
a_0a_1(t_1,\epsilon_1)\dots a_{j-1}(t_{j-1},\epsilon_{j-1})
a_j(t_{j-1})\ldots a_{n}(t_{n-1},\epsilon_{n-1})
e^{-\beta Q^2})dt d\epsilon . \cr}\ref{\intermi}
$$
Using the fact that
$$
(a_1 a_2)(t,\epsilon)=a_1(t,\epsilon)a_2(t)+a_1(t)
a_2(t,\epsilon),\ref{\leibrule}
$$
we thus obtain
$$
\eqalign{
&\beta^{n/2}\sum_{1\leq j\leq n}\;(-1)^j\tilde\tau^{\beta}_n(a_0,a_1,
\ldots ,[Q,a_j],\ldots , a_n)=\cr
&\quad\sum_{j=0}^{n-1}(-1)^j\int_{\sigma^\beta_{n-1|n-1}}
\Tr(\Gamma^na_0a_1(t_1,\epsilon_1)\dots
(a_ja_{j+1})(t_j,\epsilon_j)\ldots a_{n}(t_{n},\epsilon_{n})
e^{-\beta Q^2})dt d\epsilon =\cr
&\qquad\qquad\beta^{(n-1)/2} (b\tau^\beta)_n(a_0,a_1,\ldots ,a_n),\cr
}
$$
and the claim follows. $\square$

\section\pathintsec{Path Integral Representation of the Index}

\newsubsec
We now assume that the $\Theta$-summable module $(\h, Q, \a)$ is
associated with a supersymmetric quantum theory. This theory may
involve finitely many degrees of freedom (quantum mechanics) or
an infinite number of degrees of freedem (quantum field theory).
As a rule, a quantum theory involving finitely many degrees of
freedom leads to a $p$--summable Fredholm module with the
associated dimension equal to the number of degrees of freedom.
Quantum field theories lead to infinite dimensional
$\Theta$--summable Fredholm modules. Aside from some simple
examples, the infinite dimensional constructions described below have
not been yet carried through in a rigorous manner. Therefore, in
mathematical terms, the results formulated below have a partially
conjectural character as they rely on the existence and properties
of certain measures on infinite dimensional Grassmann algebras.

More specifically, we require the existence of a Euclidean
supersymmetric quantum field theory in the following strong sense.

\noindent
$\phantom{..}\bullet$ The algebra $\a$ consists of suitable functions
of ``time zero bosonic and fermionic field operators'' $\varphi_j(x)$
and $\psi_j(x)$, where $j = 1,\ldots, n$. Here $x\in\Sigma$, where
$\Sigma$ is a compact Riemannian manifold. In the case of quantum
mechanics, $\Sigma$ consists of a single point. We will assume that
$\varphi_j(x)$ is a real scalar field, while $\psi_j(x)$ is a
Majorana Fermi field. In fact, these are operator valued distributions,
and so only smoothed out objects $\varphi_j(f)$ and $\psi_j(f)$ (with
$f$ a test function) are well defined operators. The fact that the
number of bosonic operators is set to be equal to the number of fermionic
operators is not accidental: it reflects the underlying symmetry
(supersymmetry) of the theory. A natural way to describe a supersymmetric
theory is the language of superfields.

\noindent
$\phantom{..}\bullet$ There exists an underlying space of Euclidean
superfields. Euclidean superfields are integration variables in an
infinite dimensional Berezin integral, very much like Brownian
paths are integration variables in a Wiener integral. For our needs,
a scalar superfield $\Phi_j(x,t,\epsilon)$ ($(x,t)\in\Sigma\times
\bR$), $j = 1, \ldots , n$, has the form
$\Phi_j(x,t,\epsilon) = \phi_j(x,t)+\epsilon\Psi_j(x,t)$, where
$\phi_j(x,t)$ is a Euclidean bosonic field, and where $\Psi_j(x,t)$
is a Euclidean Majorana field. Strictly speaking, Euclidean
superfields are Grassmann algebra valued distrbutions, and so
need to be regularized to produce non-singular objects. Observe that
the algebra generated by the regularized superfields is commutative;
this is a consequence of the fact that Euclidean Bose fields commute
and Euclidean Fermi fields anticommute.

\noindent
$\phantom{..}\bullet$
The key elements of quantum theory are various Feynman - Kac formulas
which provide a bridge between the Hilbert space and Euclidean
formulations of the theory. A Feynman - Kac formula relates a
functional on the algebra $\a$ (like the vacuum value expectation,
trace or supertrace) to an integral over ${\cal M}$ with respect to
a suitable measure $d\mu(\Phi)$. For our purposes we require the
following Feynman - Kac formula. For $a_0,\ldots ,a_n\in\a$ and
$(t,\epsilon)\in\sigma^\beta_{n|n}$, there exist functions
$A_0,\ldots , A_n$ of $\Phi(.,t,\epsilon)$ such that
$$
\Tr(\Gamma a_0(t_0,\epsilon_0)a_1(t_1,\epsilon_1)\ldots
a_n(t_n,\epsilon_n) \exp(-\beta Q^2))=\int \prod_{j=0}^n
A_j(\Phi(.\;, t_j,\epsilon_j)) d\mu^\beta(\Phi),\ref{\fkfref}
$$
where $d\mu^\beta(\Phi)$ is a measure depending on $\beta$. In
concrete physical models, $\beta$ dependence means that the
measure $d\mu^\beta(\Phi)$ is concentrated on $\Phi$'s which
are periodic in $t$ with period $\beta$.

\subsec
Assuming the existence of the structure described in the previous
subsection we can prove the following theorem. It expresses the index
${\rm ind}(Q_{p+})$ of the operator $Q_p := pQp$, where $p\in\a$
is a projection, as a supersymmetric path integral. This identity
generalizes the local version of the Atiyah - Singer index theorem;
for more background and motivation see [2] and [3].
\thm\pirthm{Let P be a function of $\Phi$ corresponding to $p$.
Then,
$$
\eqalign{
{\rm ind}(Q_{p+})&=\int P(\Phi(.\;,0,0))\exp\big[-\beta^{-1}
\big(\int P(\Phi(.\; , t,\epsilon))dt d\epsilon\big)^2\big]
d\mu^\beta(\Phi)\cr
&+{1\over 2}\int\big\{1-\exp\big[-\beta^{-1}\big(\int
P(\Phi(.\; , t,\epsilon))dt d\epsilon\big)^2\big]\big\}
d\mu^\beta(\Phi).\cr}\ref{\indref}
$$
}
\medskip\noindent
{\it Proof.} According to the results of [3] and [6], the index
${\rm ind}(Q_{p+})$ can be computed from the even part of the
Chern character according to the formula
$$
{\rm ind}(Q_{p+}) = \tau^\beta_0(p)+\sum_{k=1}^\infty\;
(-1)^k{{(2k)!}\over
{k!}}\;\tau^\beta_{2k}(p-{1\over 2}, p,\ldots , p). \ref{\gsref}
$$
{}From \fkfref\ ,
$$
\eqalign{
\tau^\beta_{2k}(a_0, &a_1,\ldots , a_{2k})=\cr
&\beta^{-k}\int_{\sigma^\beta_{2k|2k}}A_0(\Phi(.\;,0,0))
A_1(\Phi(.\;,t_1,\epsilon_1))\ldots A_{2k}(\Phi(.\;,t_{2k},
\epsilon_{2k}))\; d\mu^\beta(\Phi)dt d\epsilon=\cr
&(2k)!^{-1}\beta^{-1}\int_{I^\beta_{2k|2k}}\int A(\Phi(.\;,0,0))
\prod_{j=1}^{2k}A_j(\Phi(.\;,t_j,\epsilon_j)\;
d\mu^\beta(\Phi)dtd\epsilon,\cr
}
$$
where we have used the fact that the Euclidean scalar superfields
commute with each other, and where $I^\beta_{2k|2k}$ denotes
the $(2k|2k)$-dimensional supercube of side $\beta$. Substituting
this into \gsref\ we obtain
$$
\eqalign{
{\rm ind}(Q_{p+})&=\int P(\Phi(.\;,0,0))d\mu^\beta(\Phi)\cr
&+\sum_{k=1}^\infty{{(-\beta)^k}\over{k!}}\int \big(P(\Phi(.\;,0,0))
-{1\over 2}\big)\big(\int P(\Phi(.\;,t,\epsilon))dt d\epsilon
\big)^{2k}d\mu^\beta(\Phi),\cr}
$$
and the claim follows.  $\square$

\vfill\eject

\centerline{\bf References}
\baselineskip=12pt
\frenchspacing

\bigskip

\item{1.}
Berezin, F.A.: Introduction to Superanalysis, D. Reidel Publ. Co.,
Dordrecht (1987).

\item{2.}
Connes, A.: Non-commutative differential geometry, {\it Publ. Math.
IHES}, {\bf 62} (1986), 94--144.

\item{3.}
Connes, A.: Entire cyclic cohomology of Banach algebras and
characters of $\theta$-summable Fredholm modules, {\it K-Theory},
{\bf 1} (1988), 519--548.

\item{4.}
Connes, A. and Moscovici, H.: Transgression and the Chern character
in non-commu-\break tative $K$-homology, {\it Comm. Math. Phys.},
{\bf 155} (1993), 103--122.

\item{5.}
Ernst, K., Feng, P., Jaffe, A., and Lesniewski, A.: Quantum K-theory,
II. Homotopy invariance of the Chern character, {\it J. Funct. Anal.},
{\bf 90} (1990), 355--368.

\item{6.}
Getzler, E., and Szenes, A.: On the Chern character of a theta-summable
Fredholm module, {\it J. Funct. Anal.}, {\bf 84} (1989), 343--357.

\item{7.}
Jaffe, A., Lesniewski, A., and Osterwalder, K.: Quantum K-theory,
I. The Chern character, {\it Comm. Math. Phys.}, {\bf 118} (1988),
1--14.

\item{9.}
Osterwalder, K., and Schrader, R.: Euclidean Fermi fields and a
Feynman - Kac formula for boson - fermion models, {\it Helv. Phys.
Acta}, {\bf 46} (1973), 277--302.
\vfill\eject\end